\PassOptionsToPackage{unicode}{hyperref}
\PassOptionsToPackage{hyphens}{url}
\documentclass[
]{article}
\usepackage{amsmath,amssymb}
\usepackage{lmodern}
\usepackage{iftex}
\ifPDFTeX
  \usepackage[T1]{fontenc}
  \usepackage[utf8]{inputenc}
  \usepackage{textcomp} 
\else 
  \usepackage{unicode-math}
  \defaultfontfeatures{Scale=MatchLowercase}
  \defaultfontfeatures[\rmfamily]{Ligatures=TeX,Scale=1}
\fi
\IfFileExists{upquote.sty}{\usepackage{upquote}}{}
\IfFileExists{microtype.sty}{
  \usepackage[]{microtype}
  \UseMicrotypeSet[protrusion]{basicmath} 
}{}
\makeatletter
\@ifundefined{KOMAClassName}{
  \IfFileExists{parskip.sty}{%
    \usepackage{parskip}
  }{
    \setlength{\parindent}{0pt}
    \setlength{\parskip}{6pt plus 2pt minus 1pt}}
}{
  \KOMAoptions{parskip=half}}
\makeatother
\usepackage{xcolor}
\usepackage{longtable,booktabs,array}
\usepackage{calc} 
\usepackage{etoolbox}
\makeatletter
\patchcmd\longtable{\par}{\if@noskipsec\mbox{}\fi\par}{}{}
\makeatother
\IfFileExists{footnotehyper.sty}{\usepackage{footnotehyper}}{\usepackage{footnote}}
\makesavenoteenv{longtable}
\usepackage{graphicx}
\makeatletter
\def\maxwidth{\ifdim\Gin@nat@width>\linewidth\linewidth\else\Gin@nat@width\fi}
\def\maxheight{\ifdim\Gin@nat@height>\textheight\textheight\else\Gin@nat@height\fi}
\makeatother
\setkeys{Gin}{width=\maxwidth,height=\maxheight,keepaspectratio}
\makeatletter
\def\fps@figure{htbp}
\makeatother
\ifLuaTeX
  \usepackage{selnolig}  
\fi
\IfFileExists{bookmark.sty}{\usepackage{bookmark}}{\usepackage{hyperref}}
\IfFileExists{xurl.sty}{\usepackage{xurl}}{} 
\hypersetup{
  pdftitle={ChatGPT in Veterinary Medicine: A Practical Guidance of Generative Artificial Intelligence in Clinics, Education, and Research},
  hidelinks,
  pdfcreator={LaTeX via pandoc}}

\title{ChatGPT in Veterinary Medicine: A Practical Guidance of
Generative Artificial Intelligence in Clinics, Education, and Research}
\author{}
\date{}

\begin{document}
\maketitle

Candice P. Chu\textsuperscript{1*}

Department of Veterinary Pathobiology, College of Veterinary Medicine \&
Biomedical Sciences, Texas A\&M University, College Station, TX, USA

\textbf{* Correspondence:\\
}Candice P. Chu\\
cchu@cvm.tamu.edu

Keywords: Artificial intelligence, AI, Generative AI, GenAI, Large
language model, prompt engineering, machine learning, GPT-4.

Abstract

ChatGPT, the most accessible generative artificial intelligence (AI)
tool, offers considerable potential for veterinary medicine, yet a
dedicated review of its specific applications is lacking. This review
concisely synthesizes the latest research and practical applications of
ChatGPT within the clinical, educational, and research domains of
veterinary medicine. It intends to provide specific guidance and
actionable examples of how generative AI can be directly utilized by
veterinary professionals without a programming background. For
practitioners, ChatGPT can extract patient data, generate progress
notes, and potentially assist in diagnosing complex cases. Veterinary
educators can create custom GPTs for student support, while students can
utilize ChatGPT for exam preparation. ChatGPT can aid in academic
writing tasks in research, but veterinary publishers have set specific
requirements for authors to follow. Despite its transformative
potential, careful use is essential to avoid pitfalls like
hallucination. This review addresses ethical considerations, provides
learning resources, and offers tangible examples to guide responsible
implementation. Carefully selected, up-to-date links to platforms that
host large language models are provided for advanced readers with
programming capability. A table of key takeaways was provided to
summarize this review. By highlighting potential benefits and
limitations, this review equips veterinarians, educators, and
researchers to harness the power of ChatGPT effectively.

\hypertarget{section}{%
\section{}\label{section}}

\hypertarget{introduction}{%
\section{Introduction}\label{introduction}}

Artificial intelligence (AI) is a trending topic in veterinary medicine.
Explorations in this field include applications such as dental
radiograph (1), colic detection (2), and mitosis detection in digital
pathology (3). Machine learning (ML), a subset of AI, enables systems to
learn from data without being explicitly programmed (4). Generative AI,
in turn, is a field within ML specializing in creating new content,
leading to the development of large language models (LLMs) for their
human-like text generation capabilities. Notable LLMs such as ChatGPT
(OpenAI) (5), Llama 2 (Meta) (6), Gemini (Google) (7), and Claude 2.1
(Anthropic) (8), offer diverse functionalities---some are downloadable,
others open source, or integrated into productivity software. ChatGPT
stands as the most accessible for newcomers to this technology.

GPT, or Generative Pre-trained Transformer, excels in \emph{generating}
new text, images, and other content formats rather than solely analyzing
existing data. It is \emph{pre-trained} by exposure to vast datasets of
text and code, enabling it to recognize patterns and generate human-like
responses. It employs the \emph{transformer} neural network architecture
that is particularly adept at processing language, which enables
coherent and contextually relevant outputs (9). The free version of
ChatGPT 3.5 (GPT-3.5) provides the capability of answering questions,
providing explanations, generating creative content, offering advice,
conducting research, engaging in conversation, supporting technical
tasks, aiding with education, and creating summaries. With the release
of ChatGPT 4.0 (GPT-4) (https://chat.openai.com/), a subscription-based
model, ChatGPT 4.0, the medical field has taken particular interest due
to its expanded knowledge base and capabilities in text, image, and
voice analysis and generation. Most recently, OpenAI introduced Sora, a
model capable of generating ultra-real videos solely from text prompts
(10). Despite ChatGPT's widespread use, a comprehensive review of its
applications in veterinary medicine is lacking.

The breadth of ChatGPT in medicine covers a wide range of areas, ranging
from answering patient and professional inquiries, promoting patient
engagement (11), diagnosing complex clinical cases (12), and creating
educational material (13). In veterinary medicine, ChatGPT has been
integrated to enhance virtual assistance, diagnostic accuracy,
communication with pet owners, and optimization of workflows (14--20).
While examples of ChatGPT applications are prevalent on social media and
in various publications (21--23), the best way to understand its impact
is through direct engagement. This article aims to discuss the
applications of ChatGPT in veterinary medicine, provide practical
implementations, and examine its limitations and ethical considerations.
Highlights of each section are listed in Table 1.

\textbf{ChatGPT 101: Prompts and Prompt Engineering}

Understanding prompts is crucial before engaging with ChatGPT or other
generative AI tools. Prompts act as conversation starters, consisting of
instructions or queries that elicit responses from the AI. Effective
prompts for ChatGPT integrate relevant details and context, enabling the
model to deliver precise responses (24). Prompt engineering is the
practice of refining inputs to produce optimal outputs. For instance,
researchers instructing ChatGPT to identify body condition scores from
clinical records begin prompts by detailing the data structure and
desired outcomes: ``\emph{Each row of the dataset is a different
veterinary consultation. In the column `Narrative\textquotesingle{}
there is clinical text. Your task is to extract Body Condition Score
(BCS) of the animal at the moment of the consultation if recorded. BCS
can be presented on a 9-point scale, example BCS 6/9, or on a 5-point
scale, example BCS 3.5/5. Your output should be presented in a
short-text version ONLY, following the rules below: \ldots{} (omitted)}
(24)'' Writing effective prompts involves providing contextual details
in a clear and specific way and willingness to refine them as needed.

Moreover, incorporating \textquotesingle cognitive strategy
prompts\textquotesingle{} can direct ChatGPT\textquotesingle s reasoning
more effectively (refer to Supplementary Material for more details). For
a comprehensive understanding of prompt engineering, readers are
encouraged to refer to specialized literature and open-access online
courses dedicated to this subject (25--28). Proper prompt engineering is
pivotal for shaping conversations and obtaining the intended results, as
illustrated by various examples in this review.

\textbf{Using ChatGPT in Clinical Care}

ChatGPT has the potential to provide immediate assistance upon the
client\textquotesingle s arrival at the clinic. The pre-trained
ChatGPT-4 model is adept at processing chief complaints, vital signs,
and medical histories entered by emergency medicine physicians,
subsequently making triage decisions that align closely with established
standards (29). Given that healthcare professionals in the United States
spend approximately 35\% of their time documenting patient information
(30) and that note redundancy is on the rise (31), ChatGPT
\textquotesingle s ability to distill crucial information from extensive
clinical histories and generate clinical documents are particularly
valuable (32). A veterinary study utilizing ChatGPT 3.5 Turbo for text
mining demonstrated the AI\textquotesingle s capability to pinpoint all
overweight Body Condition Score (BCS) instances within a dataset with
high precision (24). However, some limitations were noted, such as the
misclassification of lameness scoring as BCS, an issue that the
researchers believe could be addressed through refined prompt
engineering (24).

In the context of daily clinical documentation, veterinarians can input
signalment, clinical history, and physical examination findings into
ChatGPT to generate Subjective-Objective-Assessment-Plan (SOAP) notes
(30). An illustrative case presented in Supplementary Material involved
the generation of a SOAP note for a canine albuterol toxicosis incident
(33), where ChatGPT efficiently identified the diagnostic tests executed
in the case report, demonstrating that ChatGPT can be used as a
promising tool to streamline the workflow for veterinarians.

Moreover, recent research has investigated ChatGPT\textquotesingle s
proficiency in addressing clinical challenges. One study found that
GPT-4 could accurately diagnose 57\% of complex medical cases, a success
rate that outperformed 72\% of human readers of medical journals in
answering multiple-choice questions (12). Additionally,
GPT-4\textquotesingle s top diagnosis concurred with the final diagnosis
in 39\% of cases and included the final diagnosis within the top
differential diagnoses in 64\% of cases (34). With the updated image
upload function, the capability of ChatGPT-4 extends to the
interpretation of blood work images. The Supplementary Material
illustrates an example of ChatGPT analyzing Case of the Month on
eClinPath (35) and providing the correct top differential despite its
limited ability to interpret the white blood cell dot plot.

ChatGPT-4 can interpret ECGs and outperformed other LLM tools in
correctly interpreting 63\% of ECG images (36). However,
it\textquotesingle s worth noting that in a specific example provided in
the Supplementary Material, ChatGPT did not identify an atypical atrial
flutter with intermittent Ashman phenomenon in a 9-year-old Pug despite
the addition of asterisks in the ECG to indicate the wide and tall
aberrant QRS complexes (35). This example emphasizes that while ChatGPT
is a powerful tool, it cannot replace specialized AI algorithms approved
by the Food and Drug Administration (FDA) for ECG interpretation.
(37,38). Nevertheless, advances in veterinary-specific AI tools, such as
a deep learning model for canine ECG classification, are on the horizon,
with the potential to be available soon (39).

\textbf{Using ChatGPT in Veterinary Education}

Recent studies leveraging Large Language Models (LLMs) in medical
examinations underscore their utility in educational support. ChatGPT
3's performance, evaluated using 350 questions from the United States
Medical Licensing Exam (USMLE) Steps 1, 2CK, and 3, was commendable,
achieving scores near or at the passing threshold across all three
levels without specialized training (40). This evaluation involved
modifying the exam questions into various formats---open-ended or
multiple-choice with or without a forced justification---to gauge
ChatGPT's foundational medical knowledge. The AI-generated responses
often included key insights, suggesting that ChatGPT's output could
benefit medical students preparing for USMLE (40).

Another investigation benchmarked the efficacy of GPT-4, Claude 2, and
various open-source LLMs using multiple-choice questions from the
Nephrology Self-Assessment Program. Success rates varied widely, with
open-source LLMs scoring between 17.1-30.6\%, Claude 2 at 54.4\%, and
GPT-4 leading with 73.7\% (41). A comparative analysis of ChatGPT-3.5
and ChatGPT-4 indicates the newer version substantially improved in the
neonatal-perinatal medicine board examination (42). Veterinary
researchers at the University of Georgia used GPT-3.5 and GPT-4.0 to
answer faculty-generated 495 multiple-choice and true/false questions
from 15 courses in the third-year veterinary curriculum (43). The result
concurred with the previous study that GPT-4.0 (77\% correct rate)
performed substantially better than GPT-3.5 (55\% correct rate);
however, their performance is significantly lower than that of
veterinary students (86\%). These studies highlight the variances in LLM
knowledge bases, which could affect the quality of medical and
veterinary education.

Beyond exam preparation, the ChatGPT 4 Plus subscription enables users
to craft tailored versions of ChatGPT, referred to as GPTs, for diverse
uses (41). Veterinary educators, for instance, can harness these tools
to develop AI tutors to boost veterinary students\textquotesingle{}
learning. An example of a custom GPT is a specialized veterinary
clinical pathology virtual tutor named Vet Clin Path Resident (44). This
custom GPT draws from legally available open-access textbooks with
Creative Commons licenses (45--47) and the eClinPath website (35),
ensuring the information provided is sourced from credible references.
Students are encouraged to pose any question pertinent to veterinary
clinical pathology and can even request specific references or links to
web pages. More information about this GPT is detailed in the
Supplementary Material.

\textbf{Using ChatGPT in Academic Writing}

Leveraging editing services enhances clarity and minimizes grammatical
errors in scientific manuscripts, which can improve their acceptance
rate (48). While acknowledgments often thank editorial assistance, the
use of spelling-checking software is rarely disclosed. Nowadays,
AI-powered writing assistants have integrated advanced LLM capabilities
to provide nuanced suggestions for tone and context (45), thus merging
the line between original and AI-generated content. Generative AI, like
ChatGPT, extends its utility by proposing titles, structuring papers,
crafting abstracts, and summarizing research, raising questions about
the AI\textquotesingle s role in authorship as per the International
Committee of Medical Journal Editors\textquotesingle{} guidelines (49)
(Supplementary Material). Notably, traditional scientific journals are
cautious with AI, yet NEJM AI stands out for its advocacy for LLM use
(50). However, these journals still refrain from recognizing ChatGPT as
a co-author due to accountability concerns over accuracy and ethical
integrity (50--52). The academic community remains wary of
ChatGPT\textquotesingle s potential to overshadow faculty contributions
(53).

Several veterinary journals have updated their guidelines in response to
the emergence of generative AI. Among the top 20 veterinary medicine
journals as per Google Scholar (54), 14 instruct on generative AI usage
(Supplementary Material). They unanimously advise against listing AI as
a co-author, mandating disclosure of AI involvement in Methods,
Acknowledgments, or other designated sections. These recommendations
typically do not apply to basic grammar and editing tools (Supplementary
Material). Research shows ChatGPT can enhance writing efficiency,
particularly benefiting less skillful writers, suggesting
academia\textquotesingle s broader acceptance could mitigate
productivity inequality, fostering a more inclusive scholarly community
(23).

A pertinent question is the detectability of AI-generated content
without explicit disclosure. A study revealed that reviewers could
identify 68\% of ChatGPT-produced scientific abstracts; however, they
also mistakenly tagged 14\% of original works as AI-generated (55). In
another study, veterinary neurologists only had a 31-54\% success rate
in distinguishing AI-crafted abstracts (56), highlighting the risk of
misjudging authentic academic work for AI-generated content. This
misconception led to an ecologist at Cornell University being falsely
accused of writing fraud by a reviewer who deemed her work as "obviously
ChatGPT," resulting in publication rejection (57).

To counteract this, a \textquotesingle ChatGPT
detector\textquotesingle{} has been suggested. An ML tool utilizes
distinguishing features like paragraph complexity, sentence length
variability, punctuation marks, and popular wordings, achieving over
99\% effectiveness in identifying AI-authored texts (58). A subsequent
refined model can further distinguish human writings from ChatGPT-3.5
and ChatGPT-4 writings in chemistry journals with 99\% accuracy (56).
While these tools are not publicly accessible, OpenAI is developing a
classifier to flag AI-generated text (57), emphasizing the importance of
academic integrity and responsible AI use.

\textbf{ChatGPT's Ethical Issues and Limitations}

A recent survey on AI in veterinary medicine by Digitail and the
American Animal Hospital Association, involving veterinarians,
veterinary technicians/assistants, and students, showed 83.8\% of
respondents were familiar with AI and its applications in veterinary
medicine, with 69.5\% using AI tools daily or weekly (59). Yet, 36.9\%
remain skeptical, citing concerns about the systems\textquotesingle{}
reliability and accuracy (70.3\%), data security and privacy (53.9\%),
and the lack of training (42.9\%) (59). This section highlights critical
concerns to equip veterinary professionals for the upcoming AI era.

\emph{Hallucination}

Hallucination, or artificial hallucination, refers to the generation of
implausible but confident responses by ChatGPT, which poses a
significant issue (60). ChatGPT is known to create fabricated references
with incoherent Pubmed ID (61), a problem somewhat mitigated in ChatGPT
4 (18\% error rate) compared to ChatGPT 3.5 (55\% error rate) (62). The
Supplementary Material illustrated an example where ChatGPT could have
provided more accurate references, including PMIDs, underscoring its
limitations for literature searches.

\emph{Cybersecurity and Privacy}

As an LLM, ChatGPT is trained using undisclosed but purportedly
accessible online data and ongoing refinement through user interactions
during conversations (62). It raises concerns about copyright
infringement and privacy violations, as evidenced by ongoing lawsuits
against OpenAI for allegedly using private or public information without
their permission (63--65). Based on information from the OpenAI website,
user-generated content is consistently gathered and used to enhance the
service and for research purposes (66). Concerns about data security are
amplified when analyzing clinical data, suggesting a preference for
uploading de-identified datasets. Alternatively, considering local
installations of open-source, free-for-research-use LLMs, like Llama 2
or Gemma (Google), for enhanced security is recommended (67--70).

\emph{U.S. FDA Regulation}

While the FDA has approved 692 AI and ML-enabled human medical devices,
primarily in radiology (76.7\%), followed by cardiology (10.3\%) and
neurology (2.9\%) (71), veterinary medicine lacks specific premarket
requirements for AI tools. The AI- and ML-enabled veterinary products
currently span from dictation and notetaking apps (17,18), management
and communication software (19,20), radiology service (14--16), and
personalized chemotherapy (72), to name a few. These products may or may
not have scientific validation (73--80) and may be utilized by
veterinarians despite the clients\textquotesingle{} lack of consent or
complete understanding. The absence of regulatory oversight in
veterinary medicine, especially in diagnostic imaging, calls for ethical
and legal considerations to ensure patient safety in the United States
and Canada (81,82).

\emph{Practical Learning Resources}

Resources for learning about ChatGPT and generative AI are abundant,
including OpenAI's documentation (83), online courses from Vanderbilt
University via Coursera (25), Harvard University's tutorial for
generative AI (84), and the University of Michigan's guides on using
generative AI for scientific research (85). These resources are
invaluable for veterinarians seeking to navigate the evolving landscape
of AI in their practice. Last but not least, readers are advised to
engage ChatGPT with well-structured prompts, such as:
`I\textquotesingle m a veterinarian with no background in programming.
I\textquotesingle m interested in learning how to use generative AI
tools like ChatGPT. Can you recommend some resources for beginners?'
(Supplementary Material).

\textbf{The Ongoing Dialogue}

In the 2023 Responsible AI for Social and Ethical Healthcare (RAISE)
Conference held by the Department of Biomedical Informatics at Harvard
Medical School, the discussion on the judicious application of AI in
human healthcare highlighted principles that could be effectively
adapted to veterinary medicine (86). Integrating AI into veterinary
practices should amplify the benefits to animal welfare, enhance
clinical outcomes, broaden access to veterinary services, and enrich the
patient and client experience. AI should support rather than replace
veterinarians, preserving the essential human touch in animal care.

Transparent and ethical utilization of patient data is paramount,
advocating for opt-out mechanisms in data collection processes while
safeguarding client confidentiality. AI tools in the veterinary field
ought to be envisioned as adjuncts to clinical expertise, with a
potential for their role to develop progressively, subject to stringent
oversight. The growing need for direct consumer access to AI in
veterinary medicine promises advancements but necessitates meticulous
regulation to assure pet owners about data provenance and the
application of AI.

This review discussed the transformative potential of ChatGPT across
clinical, educational, and research domains within veterinary medicine.
Continuous dialogue, awareness of limitations, and regulatory oversight
are crucial to ensure generative AI augments clinical care, educational
standards, and academic ethics rather than compromising them. The
examples provided in the Supplementary Material encourage innovative
integration of AI tools into veterinary practice. By embracing
responsible adoption, veterinary professionals can harness the full
potential of ChatGPT to make the next paradigm shift in veterinary
medicine.

\hypertarget{author-contributions}{%
\section{Author Contributions}\label{author-contributions}}

All authors listed have made a substantial, direct and intellectual
contribution to the work, and approved it for publication.

\hypertarget{conflict-of-interest}{%
\section{Conflict of Interest}\label{conflict-of-interest}}

The author declares that the research was conducted in the absence of
any commercial or financial relationships that could be construed as a
potential conflict of interest.

\textbf{Data Availability}

All input prompts provided to a generative AI technology and outputs
received from a generative AI technology of this article are available
in the Supplementary Material.

\hypertarget{acknowledgment}{%
\section{Acknowledgment}\label{acknowledgment}}

ChatGPT 4.0 (\url{https://chat.openai.com/}) produced the Supplementary
Material and provided spelling, grammar, and general editing of the
original human writings.

\hypertarget{references}{%
\section{References}\label{references}}

1. Nyquist ML, Fink LA, Mauldin GE, Coffman CR. Evaluation of a novel
veterinary dental radiography artificial intelligence software program.
\emph{J Vet Dent} (2024)8987564231221071. doi: 10.1177/08987564231221071

2. Eerdekens A, Papas M, Damiaans B, Martens L, Govaere J, Joseph W,
Deruyck M. Automatic early detection of induced colic in horses using
accelerometer devices. \emph{Equine Vet J} (2024) doi: 10.1111/evj.14069

3. Rai T, Morisi A, Bacci B, Bacon NJ, Dark MJ, Aboellail T, Thomas SA,
La Ragione RM, Wells K. Keeping pathologists in the loop and an adaptive
F1-score threshold method for mitosis detection in canine perivascular
wall tumours. \emph{Cancers} (2024) 16:644. doi: 10.3390/cancers16030644

4. Rahmani AM, Yousefpoor E, Yousefpoor MS, Mehmood Z, Haider A,
Hosseinzadeh M, Ali Naqvi R. Machine learning (ML) in medicine: Review,
applications, and challenges. \emph{Mathematics} (2021) 9:2970. doi:
10.3390/math9222970

5. OpenAI. GPT-4. (2023) https://openai.com/research/gpt-4 {[}Accessed
February 17, 2024{]}

6. Touvron H, Martin L, Stone K, Albert P, Almahairi A, Babaei Y,
Bashlykov N, Batra S, Bhargava P, Bhosale S, et al. Llama 2: Open
foundation and fine-tuned chat models. (2023) doi:
10.48550/arXiv.2307.09288

7. Gemini Team, Anil R, Borgeaud S, Wu Y, Alayrac J-B, Yu J, Soricut R,
Schalkwyk J, Dai AM, Hauth A, et al. Gemini: A family of highly capable
multimodal models. (2023) doi: 10.48550/arXiv.2312.11805

8. Anthropic. Introducing Claude 2.1. (2023)
https://www.anthropic.com/news/claude-2-1 {[}Accessed February 22,
2024{]}

9. Kocoń J, Cichecki I, Kaszyca O, Kochanek M, Szydło D, Baran J,
Bielaniewicz J, Gruza M, Janz A, Kanclerz K, et al. ChatGPT: Jack of all
trades, master of none. \emph{Inf Fusion} (2023) 99:101861. doi:
10.1016/j.inffus.2023.101861

10. OpenAI. Video generation models as world simulators. (2024)
https://openai.com/research/video-generation-models-as-world-simulators
{[}Accessed February 18, 2024{]}

11. Mirza FN, Tang OY, Connolly ID, Abdulrazeq HA, Lim RK, Roye GD,
Priebe C, Chandler C, Libby TJ, Groff MW, et al. Using ChatGPT to
facilitate truly informed medical consent. \emph{NEJM AI} (2024)
1:AIcs2300145. doi: 10.1056/AIcs2300145

12. Eriksen AV, Möller S, Ryg J. Use of GPT-4 to diagnose complex
clinical cases. \emph{NEJM AI} (2023) 1:AIp2300031. doi:
10.1056/AIp2300031

13. Boscardin CK, Gin B, Golde PB, Hauer KE. ChatGPT and generative
artificial intelligence for medical education: Potential impact and
opportunity. \emph{Acad Med} (2024) 99:22. doi:
10.1097/ACM.0000000000005439

14. MetronMind. Metron IQ: Veterinary AI radiology report software.
(2023) https://www.metronmind.com/ {[}Accessed February 17, 2024{]}

15. PicoxIA. AI designed for veterinarians. (2022)
https://www.picoxia.com/en/\#scientific-validation {[}Accessed February
17, 2024{]}

16. Vetology. Veterinary radiology and artificial intelligence. (2023)
https://vetology.ai/ {[}Accessed February 17, 2024{]}

17. Dragon Veterinary. Dragon veterinary. (2020)
https://www.dragonveterinary.com/ {[}Accessed February 17, 2024{]}

18. ScribbleVet. AI for busy veterinarians. (2023)
https://www.scribblevet.com/ {[}Accessed February 17, 2024{]}

19. PetsApp. Veterinary engagement and communication platform \textbar{}
PetsApp. (2020) https://petsapp.com/ {[}Accessed February 17, 2024{]}

20. Digitail. All-in-one veterinary practice management software. (2023)
https://digitail.com/ {[}Accessed February 17, 2024{]}

21. Lee P, Goldberg C, Kohane I. \emph{The AI Revolution in Medicine:
GPT-4 and Beyond}. Pearson (2023). 289 p.

22. Taecharungroj V. ``What can ChatGPT do?'' analyzing early reactions
to the innovative AI chatbot on Twitter. \emph{Big Data Cogn Comput}
(2023) 7:35. doi: 10.3390/bdcc7010035

23. Noy S, Zhang W. Experimental Evidence on the Productivity Effects of
Generative Artificial Intelligence. (2023) doi: 10.2139/ssrn.4375283

24. Fins IS, Davies H, Farrell S, Torres JR, Pinchbeck G, Radford AD,
Noble P-J. Evaluating ChatGPT text mining of clinical records for
companion animal obesity monitoring. \emph{Vet Rec} (2024) 194:e3669.
doi: 10.1002/vetr.3669

25. White, Jules. Prompt engineering for ChatGPT. (2024)
https://www.coursera.org/learn/prompt-engineering {[}Accessed February
14, 2024{]}

26. DAIR.AI. Prompt engineering guide. (2024)
https://www.promptingguide.ai/ {[}Accessed February 19, 2024{]}

27. Ekin S. Prompt engineering for ChatGPT: A quick guide to techniques,
tips, and best practices. (2023) doi: 10.36227/techrxiv.22683919.v2

28. Akın FK. Awesome ChatGPT prompts. (2024)
https://github.com/f/awesome-chatgpt-prompts {[}Accessed February 19,
2024{]}

29. Paslı S, Şahin AS, Beşer MF, Topçuoğlu H, Yadigaroğlu M, İmamoğlu M.
Assessing the precision of artificial intelligence in emergency
department triage decisions: Insights from a study with ChatGPT.
\emph{Am J Emerg Med} (2024) 78:170--175. doi:
10.1016/j.ajem.2024.01.037

30. Nguyen J, Pepping CA. The application of ChatGPT in healthcare
progress notes: A commentary from a clinical and research perspective.
\emph{Clin Transl Med} (2023) 13:e1324. doi: 10.1002/ctm2.1324

31. Rule A, Bedrick S, Chiang MF, Hribar MR. Length and redundancy of
outpatient progress notes across a decade at an academic medical center.
\emph{JAMA Netw Open} (2021) 4:e2115334. doi:
10.1001/jamanetworkopen.2021.15334

32. Tierney AA, Gayre G, Hoberman B, Mattern B, Ballesca M, Kipnis P,
Liu V, Lee K. Ambient Artificial Intelligence Scribes to Alleviate the
Burden of Clinical Documentation. \emph{NEJM Catal} (2024)
5:CAT.23.0404. doi: 10.1056/CAT.23.0404

33. Guida SJ, Bazzle L. Rebound hyperkalemia in a dog with albuterol
toxicosis after cessation of potassium supplementation. \emph{J Vet
Emerg Crit Care} (2023) 33:715--721. doi: 10.1111/vec.13352

34. Kanjee Z, Crowe B, Rodman A. Accuracy of a generative artificial
intelligence model in a complex diagnostic challenge. \emph{JAMA} (2023)
330:78--80. doi: 10.1001/jama.2023.8288

35. Stokol, Tracy. eClinPath \textbar{} A Resource for Veterinary
Clinical Pathology. (2024) https://eclinpath.com/ {[}Accessed February
22, 2024{]}

36. Fijačko N, Prosen G, Abella BS, Metličar Š, Štiglic G. Can novel
multimodal chatbots such as Bing Chat Enterprise, ChatGPT-4 Pro, and
Google Bard correctly interpret electrocardiogram images?
\emph{Resuscitation} (2023) 193: doi:
10.1016/j.resuscitation.2023.110009

37. Alivecor. FDA clears first of its kind algorithm suite for personal
ECG. (2020)
https://alivecor.com/press/press\_release/fda-clears-first-of-its-kind-algorithm-suite-for-personal-ecg
{[}Accessed February 20, 2024{]}

38. Yao X, Rushlow DR, Inselman JW, McCoy RG, Thacher TD, Behnken EM,
Bernard ME, Rosas SL, Akfaly A, Misra A, et al. Artificial
intelligence--enabled electrocardiograms for identification of patients
with low ejection fraction: a pragmatic, randomized clinical trial.
\emph{Nat Med} (2021) 27:815--819. doi: 10.1038/s41591-021-01335-4

39. Dourson A, Santilli R, Marchesotti F, Schneiderman J, Stiel OR,
Junior F, Fitzke M, Sithirangathan N, Walleser E, Qiao X, et al.
PulseNet: Deep Learning ECG-signal classification using random
augmentation policy and continous wavelet transform for canines. (2023)
doi: 10.48550/arXiv.2305.15424

40. Kung TH, Cheatham M, Medenilla A, Sillos C, Leon LD, Elepaño C,
Madriaga M, Aggabao R, Diaz-Candido G, Maningo J, et al. Performance of
ChatGPT on USMLE: Potential for AI-assisted medical education using
large language models. \emph{PLOS Digit Health} (2023) 2:e0000198. doi:
10.1371/journal.pdig.0000198

41. Wu S, Koo M, Blum L, Black A, Kao L, Fei Z, Scalzo F, Kurtz I.
Benchmarking open-source large language models, GPT-4 and CLAUDE2 on
multiple-choice questions in nephrology. \emph{NEJM AI} (2024)
1:AIdbp2300092. doi: 10.1056/AIdbp2300092

42. Sharma P, Luo G, Wang C, Brodsky D, Martin CR, Beam A, Beam K.
Assessment of the clinical knowledge of ChatGPT-4 in neonatal-perinatal
medicine: a comparative analysis with ChatGPT-3.5. \emph{J Perinatol}
(2024)1--2. doi: 10.1038/s41372-024-01912-8

43. Coleman MC, Moore JN. Two artificial intelligence models
underperform on examinations in a veterinary curriculum. \emph{J Am Vet
Med Assoc} (2024)1--6. doi: 10.2460/javma.23.12.0666

44. Chu, Candice P. Vet Clin Path Resident: Your veterinary clinical
pathology AI tutor.
https://chat.openai.com/g/g-rfB5cBZ6X-vet-clin-path-resident {[}Accessed
February 25, 2024{]}

45. Baum, Hermann. \emph{The Lymphatic System of the Dog}. University of
Saskatchewan (1918). 212 p.
https://openpress.usask.ca/k9lymphaticsystem/ {[}Accessed February 21,
2024{]}

46. Burton, Erin. \emph{Clinical Veterinary Diagnostic Laboratory}.
University of Minnesota (2021). 287 p. https://pressbooks.umn.edu/cvdl/
{[}Accessed February 21, 2024{]}

47. Jennings, Ryan, Premanandan, Christopher. \emph{Veterinary
Histology}. The Ohio State University (2017). 245 p.
https://ohiostate.pressbooks.pub/vethisto/ {[}Accessed February 21,
2024{]}

48. Warren W. Beating the odds for journal acceptance. \emph{Sci Adv}
(2022) 8:eadd9147. doi: 10.1126/sciadv.add9147

49. ICMJE. Recommendations for the conduct, reporting, editing, and
publication of scholarly work in medical journals. (2024)
https://www.icmje.org/recommendations/ {[}Accessed February 19, 2024{]}

50. Koller D, Beam A, Manrai A, Ashley E, Liu X, Gichoya J, Holmes C,
Zou J, Dagan N, Wong TY, et al. Why we support and encourage the use of
large language models in nejm AI submissions. \emph{NEJM AI} (2023)
1:AIe2300128. doi: 10.1056/AIe2300128

51. Thorp HH. ChatGPT is fun, but not an author. \emph{Science} (2023)
379:313--313. doi: 10.1126/science.adg7879

52. Stokel-Walker C. ChatGPT listed as author on research papers: many
scientists disapprove. \emph{Nature} (2023) 613:620--621. doi:
10.1038/d41586-023-00107-z

53. Chrisinger, Ben. It's not just our students --- ChatGPT is coming
for faculty writing. \emph{Chron High Educ} (2023)
https://www.chronicle.com/article/its-not-just-our-students-ai-is-coming-for-faculty-writing
{[}Accessed February 22, 2024{]}

54. Google. Veterinary medicine: Google scholar metrics. (2024)
https://scholar.google.com/citations?view\_op=top\_venues\&hl=en\&vq=med\_veterinarymedicine
{[}Accessed February 19, 2024{]}

55. Gao CA, Howard FM, Markov NS, Dyer EC, Ramesh S, Luo Y, Pearson AT.
Comparing scientific abstracts generated by ChatGPT to real abstracts
with detectors and blinded human reviewers. \emph{Npj Digit Med} (2023)
6:1--5. doi: 10.1038/s41746-023-00819-6

56. Abani S, Volk HA, De Decker S, Fenn J, Rusbridge C, Charalambous M,
Goncalves R, Gutierrez-Quintana R, Loderstedt S, Flegel T, et al.
ChatGPT and scientific papers in veterinary neurology; is the genie out
of the bottle? \emph{Front Vet Sci} (2023) 10:
https://www.frontiersin.org/articles/10.3389/fvets.2023.1272755
{[}Accessed February 7, 2024{]}

57. Wolkovich EM. `Obviously ChatGPT' --- how reviewers accused me of
scientific fraud. \emph{Nature} (2024) doi: 10.1038/d41586-024-00349-5

58. Desaire H, Chua AE, Isom M, Jarosova R, Hua D. Distinguishing
academic science writing from humans or ChatGPT with over 99\% accuracy
using off-the-shelf machine learning tools. \emph{Cell Rep Phys Sci}
(2023) 4:101426. doi: 10.1016/j.xcrp.2023.101426

59. Danylenko, Galyna. AI in veterinary medicine: The next paradigm
shift. (2024)
https://digitail.com/blog/artificial-intelligence-in-veterinary-medicine-the-next-paradigm-shift/
{[}Accessed February 16, 2024{]}

60. Ji Z, Lee N, Frieske R, Yu T, Su D, Xu Y, Ishii E, Bang YJ, Madotto
A, Fung P. Survey of hallucination in natural language generation.
\emph{ACM Comput Surv} (2023) 55:248:1-248:38. doi: 10.1145/3571730

61. Alkaissi H, McFarlane SI. Artificial hallucinations in ChatGPT:
Implications in scientific writing. \emph{Cureus} 15:e35179. doi:
10.7759/cureus.35179

62. OpenAI. What is ChatGPT? (2024)
https://help.openai.com/en/articles/6783457-what-is-chatgpt {[}Accessed
February 17, 2024{]}

63. Kahveci ZÜ. Attribution problem of~generative AI: a view from US
copyright law. \emph{J Intellect Prop Law Pract} (2023) 18:796--807.
doi: 10.1093/jiplp/jpad076

64. Grynbaum MM, Mac R. The Times sues OpenAI and Microsoft over A.I.
use of copyrighted work. \emph{N Y Times} (2023)
https://www.nytimes.com/2023/12/27/business/media/new-york-times-open-ai-microsoft-lawsuit.html
{[}Accessed February 17, 2024{]}

65. The Authors Guild. The Authors Guild, John Grisham, Jodi Picoult,
David Baldacci, George R.R. Martin, and 13 other authors file
class-action suit against OpenAI. (2023)
https://authorsguild.org/news/ag-and-authors-file-class-action-suit-against-openai/
{[}Accessed February 17, 2024{]}

66. OpenAI. Privacy policy. (2023)
https://openai.com/policies/privacy-policy {[}Accessed February 17,
2024{]}

67. Hugging Face. Gemma release. (2024)
https://huggingface.co/collections/google/gemma-release-65d5efbccdbb8c4202ec078b
{[}Accessed February 22, 2024{]}

68. Hugging Face. Llama 2. (2024) https://huggingface.co/meta-llama
{[}Accessed February 17, 2024{]}

69. Meta. Llama. (2023) https://llama.meta.com/ {[}Accessed February 17,
2024{]}

70. Banks, Jeanine, Warkentin, Tris. Gemma: Introducing new
state-of-the-art open models. (2024)
https://blog.google/technology/developers/gemma-open-models/ {[}Accessed
February 22, 2024{]}

71. U.S. Food and Drug Administration. Artificial intelligence and
machine learning (AI/ML)-enabled medical devices. (2023)
https://www.fda.gov/medical-devices/software-medical-device-samd/artificial-intelligence-and-machine-learning-aiml-enabled-medical-devices
{[}Accessed February 17, 2024{]}

72. ImpriMed. Find the best drugs for treating your dog's lymphoma
BEFORE treatment begins. (2024) https://www.imprimedicine.com/
{[}Accessed February 25, 2024{]}

73. Müller TR, Solano M, Tsunemi MH. Accuracy of artificial intelligence
software for the detection of confirmed pleural effusion in thoracic
radiographs in dogs. \emph{Vet Radiol Ultrasound} (2022) 63:573--579.
doi: 10.1111/vru.13089

74. Boissady E, de La Comble A, Zhu X, Hespel A-M. Artificial
intelligence evaluating primary thoracic lesions has an overall lower
error rate compared to veterinarians or veterinarians in conjunction
with the artificial intelligence. \emph{Vet Radiol Ultrasound} (2020)
61:619--627. doi: 10.1111/vru.12912

75. Boissady E, De La Comble A, Zhu X, Abbott J, Adrien-Maxence H.
Comparison of a deep learning algorithm vs. humans for vertebral heart
scale measurements in cats and dogs shows a high degree of agreement
among readers. \emph{Front Vet Sci} (2021) 8:
https://www.frontiersin.org/articles/10.3389/fvets.2021.764570
{[}Accessed February 17, 2024{]}

76. Adrien-Maxence H, Emilie B, Alois DLC, Michelle A, Kate A, Mylene A,
David B, Marie de S, Jason F, Eric G, et al. Comparison of error rates
between four pretrained DenseNet convolutional neural network models and
13 board-certified veterinary radiologists when evaluating 15 labels of
canine thoracic radiographs. \emph{Vet Radiol Ultrasound} (2022)
63:456--468. doi: 10.1111/vru.13069

77. Kim E, Fischetti AJ, Sreetharan P, Weltman JG, Fox PR. Comparison of
artificial intelligence to the veterinary radiologist's diagnosis of
canine cardiogenic pulmonary edema. \emph{Vet Radiol Ultrasound} (2022)
63:292--297. doi: 10.1111/vru.13062

78. Bohannan Z, Pudupakam RS, Koo J, Horwitz H, Tsang J, Polley A, Han
EJ, Fernandez E, Park S, Swartzfager D. Predicting likelihood of in vivo
chemotherapy response in canine lymphoma using ex vivo drug sensitivity
and immunophenotyping data in a machine learning model. \emph{Vet Comp
Oncol} (2021) 19:160--171.

79. Koo J, Choi K, Lee P, Polley A, Pudupakam RS, Tsang J, Fernandez E,
Han EJ, Park S, Swartzfager D, et al. Predicting Dynamic Clinical
Outcomes of the Chemotherapy for Canine Lymphoma Patients Using a
Machine Learning Model. \emph{Vet Sci} (2021) 8:301. doi:
10.3390/vetsci8120301

80. Callegari AJ, Tsang J, Park S, Swartzfager D, Kapoor S, Choy K, Lim
S. Multimodal machine learning models identify chemotherapy drugs with
prospective clinical efficacy in dogs with relapsed B-cell lymphoma.
\emph{Front Oncol} (2024) 14:1304144. doi: 10.3389/fonc.2024.1304144

81. Bellamy JEC. Artificial intelligence in veterinary medicine requires
regulation. \emph{Can Vet J} 64:968--970.

82. Cohen EB, Gordon IK. First, do no harm. Ethical and legal issues of
artificial intelligence and machine learning in veterinary radiology and
radiation oncology. \emph{Vet Radiol Ultrasound} (2022) 63:840--850.
doi: 10.1111/vru.13171

83. OpenAI. OpenAI help center. (2024) https://help.openai.com/en/
{[}Accessed February 17, 2024{]}

84. Harvard University Information Technology. Generative artificial
intelligence (AI). (2024) https://huit.harvard.edu/ai {[}Accessed
February 17, 2024{]}

85. Michigan Institute for Data Science. A quick guide of using GAI for
scientific research. (2024)
https://midas.umich.edu/generative-ai-user-guide/ {[}Accessed February
17, 2024{]}

86. Goldberg CB, Adams L, Blumenthal D, Brennan PF, Brown N, Butte AJ,
Cheatham M, deBronkart D, Dixon J, Drazen J, et al. To Do No Harm ---
and the Most Good --- with AI in Health Care. \emph{NEJM AI} (2024)
1:AIp2400036. doi: 10.1056/AIp2400036

\hypertarget{section-1}{%
\subsection{\texorpdfstring{\hfill\break
}{ }}\label{section-1}}

\hypertarget{tables}{%
\subsection{Tables}\label{tables}}

\textbf{Table 1. Key Takeaways of the Review}

\begin{longtable}[]{@{}
  >{\raggedright\arraybackslash}p{(\columnwidth - 0\tabcolsep) * \real{1.0000}}@{}}
\toprule()
\begin{minipage}[b]{\linewidth}\raggedright
Introduction

\begin{itemize}
\item
  Machine learning (ML) is a subset of artificial intelligence (AI) that
  enables systems to learn from data without being explicitly
  programmed.
\item
  Generative AI, in turn, is a field within ML specializing in creating
  new content.
\item
  Large language models (LLMs) have human-like text generation
  capabilities. Examples include ChatGPT (OpenAI), Llama 2 (Meta),
  Gemini (Google), Gemma (Google), and Claude 2.1 (Anthropic).
\item
  GPT stands for Generative Pre-trained Transformer, indicating its
  characteristics of content generation, pre-trained by text and codes,
  and the use of transformer neural network.
\end{itemize}
\end{minipage} \\
\midrule()
\endhead
\begin{minipage}[t]{\linewidth}\raggedright
ChatGPT 101: Prompts and Prompt Engineering

\begin{itemize}
\item
  Prompts act as conversation starters, consisting of instructions or
  queries that elicit responses from the AI.
\item
  Prompt engineering is the practice of refining inputs to produce
  optimal outputs. Common strategies include providing relevant context,
  detailing the data structure, and specifying desired outcomes.
\item
  Cognitive strategy prompts can direct ChatGPT\textquotesingle s
  reasoning more effectively.
\end{itemize}
\end{minipage} \\
\begin{minipage}[t]{\linewidth}\raggedright
Using ChatGPT in Clinical Care

\begin{itemize}
\item
  ChatGPT can make triage decisions, mine text from clinical history,
  create SOAP notes, diagnose complex cases, and interpret image inputs
  such as blood work and ECG.
\end{itemize}
\end{minipage} \\
\begin{minipage}[t]{\linewidth}\raggedright
Using ChatGPT in Veterinary Education

\begin{itemize}
\item
  ChatGPT has the potential to assist medical exam takers, while the
  performance in standardized exams may vary among different LLMs.
\item
  GPTs are customized ChatGPT that can serve as an AI tutor for
  veterinary students.
\item
  Readers are encouraged to engage with Vet Clin Path Resident, a
  specialized veterinary clinical pathology virtual tutor:
  https://chat.openai.com/g/g-rfB5cBZ6X-vet-clin-path-resident
\end{itemize}
\end{minipage} \\
\begin{minipage}[t]{\linewidth}\raggedright
Using ChatGPT in Academic Writing

\begin{itemize}
\item
  Most journal publishers agree that ChatGPT cannot be listed as a
  co-author.
\item
  Several veterinary journals request authors to declare the use of
  ChatGPT in methods, acknowledgment, or designated sections in the
  manuscript.
\item
  Reviewers could mistakenly classify human writings as AI-generated
  content, while ML tools built based on specific language features
  could achieve 99\% accuracy in identifying AI-authored texts.
\item
  The official `ChatGPT detectors' are currently underdeveloped by
  OpenAI.
\end{itemize}
\end{minipage} \\
\begin{minipage}[t]{\linewidth}\raggedright
ChatGPT's Ethical Issues and Limitations

\begin{itemize}
\item
  Most veterinary professionals are familiar with AI and its application
  in veterinary medicine, while some remain skeptical about its
  reliability and accuracy, data security and privacy, and a lack of
  training.
\end{itemize}

\emph{Hallucination}

\begin{itemize}
\item
  Hallucination, or artificial hallucination, refers to the generation
  of implausible but confident responses by ChatGPT, which poses a
  significant issue.
\end{itemize}

\emph{Cybersecurity and Privacy}

\begin{itemize}
\item
  ChatGPT is trained using undisclosed but purportedly accessible online
  data, and user-generated content is consistently gathered by OpenAI.
\item
  When analyzing clinical data, uploading de-identified datasets is
  recommended.
\item
  Alternatively, considering local installations of open-source,
  free-for-research-use LLMs, like Llama 2 or Gemma, for enhanced
  security.
\end{itemize}

\emph{U.S. FDA Regulation}

\begin{itemize}
\item
  Most FDA-approved AI and ML-enabled human medical devices are in the
  field of radiology, followed by cardiovascular and neurology.
\item
  FDA has not set premarket requirements for AI tools in veterinary
  medicine.
\item
  The AI- and ML-enabled veterinary products include dictation and
  notetaking apps, management and communication software, and radiology
  services, which may or may not have scientific validation.
\end{itemize}

\emph{Practical Learning Resources}

\begin{itemize}
\item
  Resources for learning about ChatGPT and generative AI are abundant,
  including OpenAI's documentation, online courses from Vanderbilt
  University via Coursera, Harvard University's tutorial for generative
  AI, and the University of Michigan's guides on using generative AI for
  scientific research.
\item
  Readers are encouraged to ask ChatGPT about learning resources:
  https://chat.openai.com
\end{itemize}
\end{minipage} \\
\bottomrule()
\end{longtable}

\textbf{Figure Legends}

\begin{figure}
    \centering
    \includegraphics[width=1\linewidth]{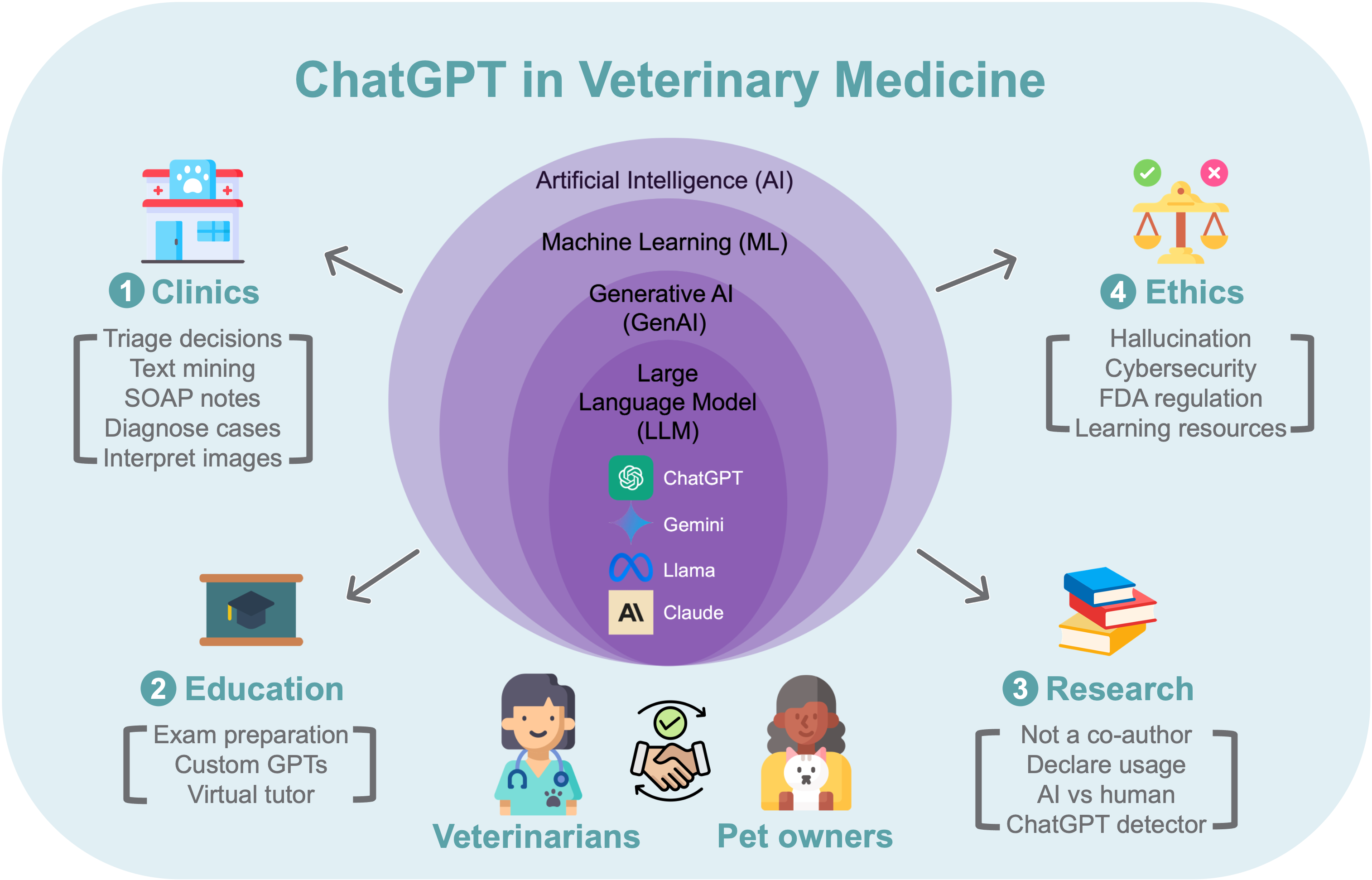}

\end{figure}

\textbf{Figure 1. Visual Abstract of the Review.}

\end{document}


Supplementary Material

All input prompts provided to a generative AI technology and outputs
received from a generative AI technology of this article are available
in this supplementary material. Results are presented in the order they
appear in the article with the section subtitle provided.

\textbf{Prompts and Prompt Engineering}

\emph{Common cognitive strategy prompts}

\begin{itemize}
\item
  Take a deep breath and think about your answers step-by-step.
\item
  Do not respond if you are unsure of the answer.
\item
  Ask yes/no questions when possible to limit responses.
\item
  Reference sources when necessary.
\item
  Ask follow-up questions if you do not understand a task.
\item
  Please do an amazing job as my career is on the line.
\end{itemize}

\emph{Example without cognitive strategy prompts}

\textbf{Prompt}:

How is diabetes managed in cats?

\textbf{ChatGPT}:

Managing diabetes in cats involves a combination of medical treatment,
dietary changes, and lifestyle adjustments to help control blood glucose
levels and minimize symptoms. The primary goals are to eliminate the
symptoms of diabetes, such as excessive thirst and urination, and to
prevent complications. Here are the key components of managing feline
diabetes:

\begin{enumerate}
\def\labelenumi{\arabic{enumi}.}
\item
  \begin{quote}
  \textbf{Insulin Therapy}: Most diabetic cats require insulin
  injections to help their bodies use glucose effectively. The type and
  dosage of insulin are determined by the cat\textquotesingle s
  individual needs and are carefully monitored and adjusted based on
  blood glucose levels.
  \end{quote}
\item
  \begin{quote}
  \textbf{Monitoring}: Regular monitoring of blood glucose levels is
  crucial in managing feline diabetes. This can involve both home
  monitoring and visits to the veterinarian. Home monitoring can include
  checking blood glucose levels with a glucometer and observing the
  cat\textquotesingle s general behavior and health.
  \end{quote}
\item
  \begin{quote}
  \textbf{Dietary Management}: A high-protein, low-carbohydrate diet is
  often recommended for diabetic cats. This type of diet can help to
  regulate blood glucose levels. Your veterinarian may recommend
  specific therapeutic diets formulated for diabetic cats.
  \end{quote}
\item
  \begin{quote}
  \textbf{Weight Management}: Maintaining an optimal body weight is
  important in managing diabetes. Overweight cats are more resistant to
  insulin, so weight loss can improve the body\textquotesingle s use of
  insulin. A balanced diet and regular exercise are key components of
  weight management.
  \end{quote}
\item
  \begin{quote}
  \textbf{Regular Exercise}: Regular physical activity can help to
  improve insulin sensitivity and overall health. Encouraging play and
  providing opportunities for exercise can be beneficial for diabetic
  cats.
  \end{quote}
\item
  \begin{quote}
  \textbf{Routine Veterinary Check-ups}: Regular check-ups with a
  veterinarian are essential to monitor the cat\textquotesingle s
  diabetes and overall health. These visits may include blood tests,
  urine tests, and assessments of the cat\textquotesingle s response to
  treatment.
  \end{quote}
\item
  \begin{quote}
  \textbf{Management of Concurrent Illnesses}: It\textquotesingle s
  important to manage any other health conditions the cat may have, as
  these can impact diabetes management. Conditions such as dental
  disease, urinary tract infections, or kidney disease can affect blood
  glucose levels and the overall health of the cat.
  \end{quote}
\item
  \begin{quote}
  \textbf{Education and Support}: Caring for a diabetic cat requires a
  commitment from the cat owner. Veterinary teams often provide
  education and support to help owners understand the disease, treatment
  options, and how to manage their cat\textquotesingle s condition at
  home.
  \end{quote}
\end{enumerate}

It\textquotesingle s important to work closely with a veterinarian to
develop and adjust a management plan that\textquotesingle s tailored to
the individual needs of the diabetic cat. With proper management, many
diabetic cats can lead happy, active lives.

\emph{Example with cognitive strategy prompts}

\textbf{Prompt}:

Could you provide a detailed, step-by-step guide on managing diabetes in
cats, including dietary recommendations, insulin therapy, and
monitoring? I am a veterinarian. It\textquotesingle s crucial to get
this right for the well-being of the feline patient, so please reference
up-to-date veterinary guidelines.

\textbf{ChatGPT}:

Managing diabetes in cats requires a comprehensive approach that
includes dietary management, insulin therapy, and regular monitoring of
blood glucose levels. Here\textquotesingle s a detailed, step-by-step
guide based on current veterinary practices and guidelines:

\textbf{1. Diagnosis and Initial Evaluation}

\begin{itemize}
\item
  Confirm the diagnosis of diabetes mellitus through clinical signs and
  laboratory tests, including blood glucose concentration and urinalysis
  for glucose and ketones.
\item
  Evaluate the cat for concurrent diseases, such as urinary tract
  infections, chronic kidney disease, or hyperthyroidism, which can
  complicate diabetes management.
\item
  Determine the cat\textquotesingle s body condition score and weight to
  tailor dietary and insulin therapy.
\end{itemize}

\textbf{2. Dietary Management}

\begin{itemize}
\item
  Diet plays a crucial role in managing feline diabetes. The goals are
  to achieve and maintain ideal body weight and to stabilize blood
  glucose levels.
\item
  High-protein, low-carbohydrate diets are recommended for diabetic
  cats, as they help improve insulin sensitivity and control blood
  glucose levels.
\item
  Gradual weight loss is important for overweight cats. Aim for 1-2\%
  body weight loss per week to avoid hepatic lipidosis.
\item
  Consistent feeding times and amounts are essential, especially if the
  cat is on insulin. Some veterinarians recommend feeding cats at the
  time of insulin administration or shortly before.
\end{itemize}

\textbf{3. Insulin Therapy}

\begin{itemize}
\item
  Insulin therapy is often necessary for diabetic cats. The choice of
  insulin type and dose is based on the individual cat\textquotesingle s
  needs and response.
\item
  Long-acting insulin formulations (e.g., glargine, detemir) are
  commonly used in cats and are preferred for initiating therapy in
  newly diagnosed cases.
\item
  Start with a conservative insulin dose to avoid hypoglycemia. Dosing
  adjustments should be made based on blood glucose monitoring and
  clinical response.
\item
  Monitor for signs of insulin overdose, such as weakness, lethargy,
  tremors, or seizures, which require immediate veterinary attention.
\end{itemize}

\textbf{4. Monitoring}

\begin{itemize}
\item
  Home blood glucose monitoring is highly recommended for diabetic cats.
  It provides more consistent and stress-free readings than in-clinic
  measurements.
\item
  Teach owners how to perform home blood glucose testing using a
  glucometer and how to record the results.
\item
  Blood glucose curves, where glucose levels are measured every 1-2
  hours over a 12-hour period, are useful for evaluating insulin
  efficacy and deciding on dose adjustments.
\item
  Regular veterinary check-ups (every 3-6 months) are necessary to
  assess overall health, review blood glucose logs, and adjust treatment
  as needed.
\end{itemize}

\textbf{5. Management of Complications}

\begin{itemize}
\item
  Be vigilant for diabetic complications, such as diabetic ketoacidosis,
  which is a medical emergency requiring immediate veterinary care.
\item
  Regularly screen for concurrent diseases that can affect diabetes
  management, such as dental disease, pancreatitis, or infections.
\end{itemize}

\textbf{6. Owner Education}

\begin{itemize}
\item
  Educate owners about the importance of adherence to the treatment
  plan, potential complications, and the signs of hypoglycemia.
\item
  Encourage owners to maintain a consistent routine regarding feeding
  and insulin administration.
\item
  Discuss the long-term commitment and potential financial implications
  of managing a diabetic cat.
\end{itemize}

\textbf{7. Reassessment and Adjustment}

\begin{itemize}
\item
  Regularly reassess the cat\textquotesingle s clinical signs, body
  weight, and blood glucose levels to adjust the management plan as
  needed.
\item
  Some cats may experience diabetic remission, especially with early
  diagnosis and appropriate management. In such cases, insulin may be
  gradually reduced and possibly discontinued under close veterinary
  supervision.
\end{itemize}

\textbf{References}

\begin{itemize}
\item
  For up-to-date guidelines, consult the American Animal Hospital
  Association (AAHA) Diabetes Management Guidelines for Dogs and Cats
  and the International Society of Feline Medicine (ISFM) consensus
  guidelines on the practical management of diabetes mellitus in cats.
\end{itemize}

This guide combines general principles with current best practices in
veterinary medicine. However, each diabetic cat is unique, and treatment
should be tailored to the individual patient\textquotesingle s needs and
response to therapy. Continuous education and collaboration with a
veterinary endocrinologist may also be beneficial in complex
cases.\textbf{\hfill\break
}

\textbf{Using ChatGPT in Clinical Care}

\emph{SOAP note}

\textbf{Prompt:} Write a concise and accurate health progress note: A
3-year-old female neutered mixed breed dog weighing 38 kg was presented
for acute onset weakness and difficulty breathing. The dog was
clinically normal until 8 hours prior to presentation, when inappetence
and weakness became apparent. No other clinical signs were noted. There
was concern for albuterol inhalation, as the dog
walker\textquotesingle s missing albuterol inhaler was found with bite
marks present just prior to the onset of clinical signs. The dog had a
history of dexmedetomidine-induced ventricular arrhythmias, as well as
elbow dysplasia, for which it was receiving carprofen (50--75 mg, PO, q
12 h as needed). The last dose of carprofen was 5 days prior to
presentation. There was no other pertinent medical history.

\textbf{ChatGPT:}

\textbf{Subjective:}

3-year-old female neutered mixed breed dog presented with acute onset of
weakness and difficulty breathing. The patient was asymptomatic until 8
hours before presentation, showing signs of inappetence and weakness.
The owner reports a possible albuterol inhalation incident, as indicated
by the dog walker\textquotesingle s albuterol inhaler found with bite
marks just before the clinical signs appeared. The dog has a known
history of ventricular arrhythmias triggered by dexmedetomidine and
elbow dysplasia, managed intermittently with carprofen, last
administered 5 days ago.

\textbf{Objective:}

\begin{itemize}
\item
  Weight: 38 kg
\item
  Clinical signs: Acute weakness, difficulty breathing, inappetence.
\item
  Medical history: Dexmedetomidine-induced ventricular arrhythmias,
  elbow dysplasia.
\item
  Medication: Carprofen 50--75 mg, PO, q 12 h as needed (last dose 5
  days prior).
\item
  Possible albuterol inhaler ingestion/interaction.
\end{itemize}

\textbf{Assessment:}

The acute onset of symptoms, particularly the difficulty breathing and
weakness, could be indicative of albuterol toxicity, given the temporal
relationship with the inhaler incident. The patient\textquotesingle s
history of ventricular arrhythmias may complicate the presentation and
necessitates careful cardiovascular monitoring. The lack of recent
carprofen administration suggests it is less likely to be contributing
to the current clinical picture.

\textbf{Plan:}

\begin{enumerate}
\def\labelenumi{\arabic{enumi}.}
\item
  \begin{quote}
  Immediate stabilization, including oxygen therapy and IV fluid
  support, as needed based on respiratory and cardiovascular assessment.
  \end{quote}
\item
  \begin{quote}
  Comprehensive cardiovascular monitoring, considering the history of
  ventricular arrhythmias and potential for albuterol to exacerbate this
  condition.
  \end{quote}
\item
  \begin{quote}
  Bloodwork to assess baseline organ function and electrolyte
  imbalances, which may be affected by albuterol toxicity.
  \end{quote}
\item
  \begin{quote}
  Consider activated charcoal administration if within the window of
  opportunity and if deemed safe based on the patient\textquotesingle s
  respiratory status.
  \end{quote}
\item
  \begin{quote}
  Symptomatic and supportive care tailored to the
  patient\textquotesingle s evolving clinical signs and diagnostic
  findings.
  \end{quote}
\item
  \begin{quote}
  Close monitoring and reassessment of the patient\textquotesingle s
  respiratory and cardiovascular status, adjusting the treatment plan as
  necessary.
  \end{quote}
\end{enumerate}

Inform the owner of the situation, potential complications, and the need
for possible intensive care.

\emph{Blood work Interpretation}

Case source: \url{https://eclinpath.com/case-1-2024/}

Readers can compare ChatGPT's responses to the answers at
\url{https://eclinpath.com/case-1-2024/2/}

\textbf{Prompt:}

A 15-week-old female Maltese puppy weighing 1.5 kg was presented to the
primary care veterinarian with a one day history of lethargy and
vomiting. The puppy was purchased from a breeder 6 weeks prior and had
received vaccinations for Bordetella vaccine and distemper, adenovirus,
parainfluenza, parvovirus (DAPP x 2). She did not interact with other
animals or walk on the street, but did have access to a private outdoor
backyard behind the owner's apartment where she was let out daily. There
was no known history of dietary indiscretion or foreign material
ingestion. The owner declined diagnostic testing at the initial visit
and the puppy was given an anti-emetic injection and one dose of an oral
dewormer. The puppy's vomiting and lethargy persisted and she was
presented to the Emergency Service the next day for evaluation. On
physical examination, she had a dull demeanor, dry/tacky mucous
membranes, and mildly icteric skin. Other than the treatments given the
day prior by her primary care veterinarian, the puppy was not on any
medications. A point-of-care Parvovirus Fecal Antigen ELISA snap test
was negative. Abdominal radiographs were unremarkable.

This patient's complete blood count (CBC, with a point-of-care Procyte
analyzer), biochemistry results, and the white blood cell (WBC) dot plot
from the analyzer are attached. Evaluate these results in conjunction
with the history and answer the following questions:

1. Interpret the leukogram. With no blood smear available to review,
what do you think about the accuracy of the CBC differential count when
looking at the WBC dot plot?

2. What is your top differential diagnosis?

3. What are potential findings in a urinalysis from this patient as a
result of your top differential diagnosis?

\textbf{ChatGPT:}

Based on the information provided and the images of the complete blood
count (CBC) results, biochemistry results, and the white blood cell
(WBC) dot plot, here is an interpretation:

\begin{enumerate}
\def\labelenumi{\arabic{enumi}.}
\item
  \textbf{Interpretation of the leukogram and accuracy of the CBC
  differential count:}
\end{enumerate}

The CBC results from the Procyte analyzer show several abnormalities:

\begin{itemize}
\item
  Elevated white blood cells (WBCs) at 47.6 K/µL, which is above the
  normal reference interval, indicating leukocytosis.
\item
  Neutrophils are increased at 39.5 K/µL, also above the reference
  range, suggesting neutrophilic leukocytosis. This is often associated
  with inflammation, infection, or stress.
\item
  Lymphocytes are within normal limits.
\item
  Monocytes are elevated at 5.8 K/µL, which can be a response to chronic
  inflammation.
\item
  The presence of (H) next to some values indicates that the results are
  higher than the normal reference range.
\end{itemize}

The WBC dot plot is a graphical representation of cell populations based
on size and granularity. It aids in distinguishing different types of
white blood cells. Without a blood smear to manually verify the
automated counts, there is always some uncertainty about the accuracy of
the differential count, especially if there are abnormal or atypical
cells that the analyzer may not categorize correctly. In this case, the
dot plot seems to show distinct populations of eosinophils (EOS),
monocytes (MONO), reticulocytes (URBC), lymphocytes (LYM), neutrophils
(NEU), and basophils (BASO), which could imply that the analyzer is
effectively differentiating the cell types, but caution should be taken
with the high counts of monocytes and neutrophils.

\begin{enumerate}
\def\labelenumi{\arabic{enumi}.}
\setcounter{enumi}{1}
\item
  \textbf{Top differential diagnosis:}
\end{enumerate}

Considering the history of vomiting, lethargy, icteric skin, and the
laboratory results showing high WBCs, neutrophilia, monocytosis,
elevated liver enzymes (ALT), and hyperbilirubinemia (as indicated by
the icteric plasma appearance), a top differential diagnosis could be a
hepatobiliary disease such as cholangiohepatitis or leptospirosis. The
high creatinine and urea also point to kidney involvement, which is
common with leptospirosis.

\begin{enumerate}
\def\labelenumi{\arabic{enumi}.}
\setcounter{enumi}{2}
\item
  \textbf{Potential findings in a urinalysis from this patient:}
\end{enumerate}

If the top differential diagnosis is a hepatobiliary disease or
leptospirosis, urinalysis might show:

\begin{itemize}
\item
  Bilirubinuria: the presence of bilirubin in the urine which would
  correspond with the icteric plasma and suggest liver dysfunction.
\item
  Proteinuria: increased protein in the urine, which can be associated
  with kidney damage.
\item
  Hematuria: blood in the urine, which can occur with kidney involvement
  or infection.
\item
  Pyuria: white blood cells in the urine, suggesting urinary tract
  infection or inflammation.
\item
  Casts: cellular casts could indicate tubular damage if kidney
  involvement is present.
\end{itemize}

The findings of urinalysis would help further support the diagnosis by
indicating liver dysfunction and possible kidney impairment, which are
consistent with conditions like leptospirosis, especially in a young
puppy with outdoor access and no known ingestion of foreign material.
Further diagnostic tests, including serology for leptospirosis,
ultrasound, and possibly a liver biopsy, might be indicated based on
these findings.

\emph{ECG Interpretation}

\textbf{Prompt:}

User Uploaded image (Figure 1 in Chen, K., Tanner, M., \& Thomason, J.
D. (2022). ECG of the Month. Journal of the American Veterinary Medical
Association, 260(13), 1641-1643.)

A 9-year-old 10-kg spayed female Pug was presented to the emergency
service at the Kansas State University Veterinary Health Center with
chief complaints of tachypnea and dyspnea of 3 to 4 days' duration. On
presentation, the dog was quiet, alert, and responsive. Rectal
temperature was 38.7 °C; heart rate was 140 beats/min. The dog was
panting with a noticeable expiratory push. Stenotic nares, stertorous
breathing, and increased referred upper respiratory noise were also
noted. There was no heart murmur auscultated on physical examination.

Results of a CBC and serum biochemical profile were unremarkable.
Thoracic radiographic findings were consistent with aspiration pneumonia
and pectus excavatum. Echocardiography was performed to evaluate for
pulmonary arterial hypertension and revealed normal chamber sizes and an
absence of pulmonary arterial hypertension. An ECG recording was
obtained and attached. You are a board-certified veterinary
cardiologists. Please provide your interpretation.

\textbf{ChatGPT:}

The ECG you\textquotesingle ve provided seems to show a normal sinus
rhythm, characterized by the presence of P waves before each QRS
complex, and a normal QRS morphology and rate. The rate being 140 beats
per minute is somewhat elevated; however, this may be due to stress or
pain, especially considering the dog\textquotesingle s clinical signs of
respiratory distress.

Notably, there are no gross abnormalities in the QRS complexes or T
waves in the ECG tracings provided. There is no evidence of significant
arrhythmias, conduction blocks, or obvious chamber enlargement. No
significant ST-segment changes are apparent that would suggest
myocardial ischemia or injury.

However, it is essential to correlate these findings with clinical
signs, as some abnormalities may not be detected on an ECG alone. The
fact that the dog is presenting with tachypnea and dyspnea, and the ECG
appears unremarkable, suggests that the cardiac function is not the
primary issue, consistent with the echocardiographic findings that did
not show pulmonary arterial hypertension or chamber enlargement.

Given the absence of pulmonary arterial hypertension and significant ECG
changes, and considering the clinical signs and thoracic radiographic
findings, it would be reasonable to focus on the management of
aspiration pneumonia and the structural abnormality of pectus excavatum,
both of which can significantly affect respiratory function. The
stertorous breathing and increased referred upper respiratory noise
might be related to the dog\textquotesingle s breed-specific issues,
such as brachycephalic airway syndrome, which are common in Pugs and can
exacerbate respiratory difficulties.

Continued monitoring of the dog\textquotesingle s cardiac function may
be prudent, but treatment should likely address the primary respiratory
conditions and any supportive care to alleviate the
dog\textquotesingle s dyspnea and tachypnea. It\textquotesingle s always
recommended to have a face-to-face consultation with a veterinary
cardiologist for a thorough evaluation and diagnosis.

\textbf{\hfill\break
}

\textbf{Using ChatGPT in Veterinary Education}

Vet Clin Path Resident: Your veterinary clinical pathology AI tutor

(\url{https://chat.openai.com/g/g-rfB5cBZ6X-vet-clin-path-resident})

\textbf{Prompt:} What are your references? Do you have web browsing
capability?

\textbf{Vet Clin Path Resident:\\
}My references for providing educational support in veterinary clinical
pathology include the following textbooks:

\begin{enumerate}
\def\labelenumi{\arabic{enumi}.}
\item
  \textbf{Clinical Veterinary Diagnostic Laboratory}
\end{enumerate}

\begin{itemize}
\item
  This resource covers various diagnostic techniques and laboratory
  procedures used in veterinary medicine, helping to understand how
  diseases are diagnosed and monitored.
\end{itemize}

\begin{enumerate}
\def\labelenumi{\arabic{enumi}.}
\setcounter{enumi}{1}
\item
  \textbf{Veterinary Histology}
\end{enumerate}

\begin{itemize}
\item
  This reference provides detailed information on the microscopic
  structure of animal tissues, which is essential for understanding
  disease processes and pathology.
\end{itemize}

\begin{enumerate}
\def\labelenumi{\arabic{enumi}.}
\setcounter{enumi}{2}
\item
  \textbf{The Lymphatic System of the Dog}
\end{enumerate}

\begin{itemize}
\item
  This book offers in-depth knowledge about the lymphatic system in
  dogs, which is crucial for diagnosing and understanding various
  diseases, including cancers and infections that affect the lymphatic
  system.
\end{itemize}

Regarding web browsing capability, I do not have the ability to browse
the internet or access real-time information. My knowledge and resources
are limited to the content of the textbooks provided and specific
educational websites like the eClinPath website for veterinary clinical
pathology. If you have any questions or need information within these
resources, I\textquotesingle m here to help!

\textbf{Prompt:} What can you do for veterinary students who are
preparing for the mid-term exam in veterinary clinical pathology?

\textbf{Vet Clin Path Resident:\\
}As Vet Clin Path Resident, I can assist veterinary students preparing
for their mid-term exam in veterinary clinical pathology in several
ways:

\begin{enumerate}
\def\labelenumi{\arabic{enumi}.}
\item
  \textbf{Concept Clarification}: If there are specific topics or
  concepts you\textquotesingle re struggling with, I can help explain
  them. This could range from understanding basic hematologic parameters
  to more complex diseases affecting the lymphatic system.
\item
  \textbf{Case-Based Scenarios}: I can provide hypothetical case
  scenarios that you might encounter in your exam. These scenarios can
  help you apply theoretical knowledge to practical situations,
  enhancing your critical thinking and diagnostic skills.
\item
  \textbf{Review Questions}: I can create or explain review questions
  based on the content of the textbooks I have access to. This can help
  reinforce your understanding of key concepts and prepare you for the
  types of questions you might face.
\item
  \textbf{Terminology}: If you\textquotesingle re having trouble with
  specific terminology used in veterinary clinical pathology, I can
  provide definitions and explanations, helping you become more familiar
  with the language used in the field.
\item
  \textbf{Study Strategies}: While I can\textquotesingle t offer
  personalized advice, I can suggest general strategies for studying
  clinical pathology, such as focusing on understanding disease
  mechanisms rather than memorizing facts.
\item
  \textbf{Resource Summaries}: I can provide summaries of relevant
  sections from the textbooks "The Lymphatic System of the Dog",
  "Clinical Veterinary Diagnostic Laboratory", and "Veterinary
  Histology". These summaries can serve as quick reference guides to
  important topics.
\end{enumerate}

Please let me know which areas you\textquotesingle d like to focus on or
if you have specific questions or topics in mind!

\textbf{\hfill\break
}

\textbf{Using ChatGPT in Veterinary Research}

\emph{Scientific writing}

\textbf{Prompt:} How can you help in scientific writing?

\textbf{ChatGPT 4:}

I can assist with scientific writing in various ways:

\begin{enumerate}
\def\labelenumi{\arabic{enumi}.}
\item
  \textbf{Idea Generation}: I can help brainstorm ideas for research
  projects, papers, or experiments based on current trends and gaps in
  your field of study.
\item
  \textbf{Outline Creation}: I can help create an outline for your
  scientific paper, including sections such as the abstract,
  introduction, methods, results, discussion, and conclusion.
\item
  \textbf{Research and References}: While I can\textquotesingle t browse
  the internet in real-time, I can provide information up to my last
  training cut-off in December 2023, including summaries of scientific
  concepts, methodologies, and historical data. I can also guide you on
  how to structure citations and references according to various style
  guides (APA, MLA, Chicago, etc.).
\item
  \textbf{Writing Assistance}: I can assist in drafting sections of your
  paper, from complex explanations of scientific concepts to discussions
  of your results. I can also help with paraphrasing, ensuring clarity,
  and maintaining a formal tone suitable for scientific discourse.
\item
  \textbf{Data Analysis and Visualization}: I can provide basic
  assistance with statistical analysis and help generate or interpret
  data visualizations, graphs, and tables that you can use in your
  paper.
\item
  \textbf{Editing and Proofreading}: I can help review your drafts for
  grammatical errors, coherence, clarity, and adherence to scientific
  writing standards. I can also suggest improvements in sentence
  structure, vocabulary, and overall flow.
\item
  \textbf{Practice and Feedback}: I can provide questions or prompts to
  help you practice writing about your research topic, and give feedback
  on your responses to help you refine your writing skills.
\item
  \textbf{Literature Review Assistance}: I can help summarize and
  synthesize existing literature, though I\textquotesingle m limited to
  the information available up to my last training update. I can also
  suggest structures for organizing your literature review.
\item
  \textbf{Ethical Considerations and Compliance}: I can provide
  information on ethical considerations in research and writing,
  including issues related to plagiarism, data fabrication, and the
  proper attribution of sources.
\item
  \textbf{Publication Guidance}: I can offer insights into the
  publication process, including selecting appropriate journals,
  understanding the peer-review process, and responding to
  reviewers\textquotesingle{} comments.
\end{enumerate}

Please provide specific details or sections of your scientific writing
where you need assistance, and I\textquotesingle ll do my best to
support you.

\textbf{Prompt:} Which of these ten abilities would meet authorship
requirements in typical scientific journals?

\textbf{ChatGPT 4:\\
}In the context of scientific publishing, authorship criteria can vary
slightly between journals but generally adhere to standards set by
organizations such as the International Committee of Medical Journal
Editors (ICMJE) or specific discipline guidelines. According to the
ICMJE, to qualify for authorship, contributors must meet the following
four criteria:

\begin{enumerate}
\def\labelenumi{\arabic{enumi}.}
\item
  \begin{quote}
  \textbf{Substantial contributions to the conception or design of the
  work; or the acquisition, analysis, or interpretation of data for the
  work; AND}
  \end{quote}
\item
  \begin{quote}
  \textbf{Drafting the work or revising it critically for important
  intellectual content; AND}
  \end{quote}
\item
  \begin{quote}
  \textbf{Final approval of the version to be published; AND}
  \end{quote}
\item
  \begin{quote}
  \textbf{Agreement to be accountable for all aspects of the work in
  ensuring that questions related to the accuracy or integrity of any
  part of the work are appropriately investigated and resolved.}
  \end{quote}
\end{enumerate}

Based on these criteria, the following abilities from the list I
provided would typically meet authorship requirements:

\begin{itemize}
\item
  \begin{quote}
  \textbf{Idea Generation}: Contributing original ideas for the research
  project could meet the first criterion if these ideas are fundamental
  to the conception or design of the work.
  \end{quote}
\item
  \begin{quote}
  \textbf{Research and References}: Engaging in substantial research,
  especially if it involves the acquisition, analysis, or interpretation
  of data, could fulfill the first criterion.
  \end{quote}
\item
  \begin{quote}
  \textbf{Writing Assistance}: Drafting significant sections of the
  manuscript or revising it critically for important intellectual
  content aligns with the second criterion.
  \end{quote}
\item
  \begin{quote}
  \textbf{Data Analysis and Visualization}: Substantial contributions to
  the analysis of data for the work could meet the first criterion.
  \end{quote}
\item
  \begin{quote}
  \textbf{Editing and Proofreading}: This could contribute to the second
  criterion if the editing involves critical revisions for important
  intellectual content, not just grammar or style corrections.
  \end{quote}
\end{itemize}

\textbf{Literature Review Assistance}: If this involves substantial
contributions to the conception of the work or critical analysis and
synthesis of the literature that shapes the research questions or
interpretation of data, it could meet the first and potentially the
second criterion.

The other abilities I listed, such as Outline Creation, Practice and
Feedback, Ethical Considerations and Compliance, and Publication
Guidance, are supportive in nature and typically do not directly qualify
for authorship on their own. These contributions are often acknowledged
in the acknowledgments section of a paper rather than warranting
authorship, as they don\textquotesingle t directly meet the
ICMJE\textquotesingle s four criteria for authorship.

It\textquotesingle s important to note that authorship norms can vary
somewhat by discipline and journal, so it\textquotesingle s crucial to
consult the specific authorship guidelines of the journal
you\textquotesingle re considering for publication.

\textbf{Top 20 Journals in Google Scholar Metrics -- Veterinary
Medicine}

Whether the author guidelines include artificial intelligence

\begin{enumerate}
\def\labelenumi{\arabic{enumi}.}
\item
  \begin{quote}
  \textbf{Animals:} N/A
  \end{quote}
\item
  \begin{quote}
  \textbf{Frontiers in Veterinary Science}:
  \end{quote}
\end{enumerate}

\begin{quote}
These guidelines cover acceptable uses of generative AI technologies
such as Large Language Models (ChatGPT, Jasper) and text-to-image
generators (DALL-E 2, Midjourney, Stable Diffusion) in the writing or
editing of manuscripts submitted to Frontiers.

Authors should not list a generative AI technology as a co-author or
author of any submitted manuscript. Generative AI technologies cannot be
held accountable for all aspects of a manuscript and consequently do not
meet the criteria required for authorship.

If the author of a submitted manuscript has used written or visual
content produced by or edited using a generative AI technology, this use
must follow all Frontiers guidelines and policies. Specifically, the
author is responsible for checking the factual accuracy of any content
created by the generative AI technology. This includes, but is not
limited to, any quotes, citations or references. Figures produced by or
edited using a generative AI technology must be checked to ensure they
accurately reflect the data presented in the manuscript. Authors must
also check that any written or visual content produced by or edited
using a generative AI technology is free from plagiarism.

If the author of a submitted manuscript has used written or visual
content produced by or edited using a generative AI technology, such use
must be acknowledged in the acknowledgements section of the manuscript
and the methods section if applicable. This explanation must list the
name, version, model, and source of the generative AI technology.

We encourage authors to upload all input prompts provided to a
generative AI technology and outputs received from a generative AI
technology in the supplementary files for the manuscript.
\end{quote}

\begin{enumerate}
\def\labelenumi{\arabic{enumi}.}
\setcounter{enumi}{2}
\item
  \begin{quote}
  \textbf{Transboundary and Emerging Diseases}:
  \end{quote}

  You can make use of a free artificial intelligence (AI)-based language
  editing tool, Writefull, at the point of submission. Writefull will
  scan your manuscript and make suggestions to help improve the quality
  of your writing. The tool applies machine learning, trained on
  millions of published scientific articles, and suggests improvements
  to grammar, spelling, and academic language.
\item
  \begin{quote}
  \textbf{Journal of Veterinary Internal Medicine}:
  \end{quote}

  Artificial Intelligence Generated Content (AIGC) tools---such as
  ChatGPT and others based on large language models (LLMs)---cannot be
  considered capable of initiating an original piece of research without
  direction by human authors. They also cannot be accountable for a
  published work or for research design, which is a generally held
  requirement of authorship (as discussed in the previous section), nor
  do they have legal standing or the ability to hold or assign
  copyright. Therefore---in accordance with COPE's position statement on
  AI tools---these tools cannot fulfill the role of, nor be listed as,
  an author of an article. If an author has used this kind of tool to
  develop any portion of a manuscript, its use must be described,
  transparently and in detail, in the Methods section. The author is
  fully responsible for the accuracy of any information provided by the
  tool and for correctly referencing any supporting work on which that
  information depends. Tools that are used to improve spelling, grammar,
  and general editing are not included in the scope of these guidelines.
  The final decision about whether use of an AIGC tool is appropriate or
  permissible in the circumstances of a submitted manuscript or a
  published article lies with the Journal's editor or other party
  responsible for the publication's editorial policy.
\item
  \begin{quote}
  \textbf{BMC Veterinary Research}:
  \end{quote}
\end{enumerate}

\begin{quote}
Large Language Models (LLMs), such as ChatGPT, do not currently satisfy
our authorship criteria. Notably an attribution of authorship carries
with it accountability for the work, which cannot be effectively applied
to LLMs. Use of an LLM should be properly documented in the Methods
section (and if a Methods section is not available, in a suitable
alternative part) of the manuscript.

The fast moving area of generative AI image creation has resulted in
novel legal copyright and research integrity issues. As publishers, we
strictly follow existing copyright law and best practices regarding
publication ethics. While legal issues relating to AI-generated images
and videos remain broadly unresolved, Springer Nature journals are
unable to permit its use for publication.

Exceptions are images/art obtained from agencies that we have
contractual relationships with that have created images in a legally
acceptable manner. Other exceptions to this policy include images and
video that are directly referenced in a piece that is specifically about
AI and will be reviewed on a case-by-case basis.

As we expect things to develop rapidly in this field in the near future,
we will review this policy regularly and adapt it if necessary.

Please note: Not all AI tools are generative. The use of non-generative
machine learning tools to manipulate, combine or enhance existing images
or figures should be disclosed in the relevant caption upon submission
to allow a case-by-case review.
\end{quote}

\begin{enumerate}
\def\labelenumi{\arabic{enumi}.}
\setcounter{enumi}{5}
\item
  \begin{quote}
  \textbf{Veterinary Microbiology}:
  \end{quote}
\end{enumerate}

\begin{quote}
Where authors use generative artificial intelligence (AI) and
AI-assisted technologies in the writing process, authors should only use
these technologies to improve readability and language. Applying the
technology should be done with human oversight and control, and authors
should carefully review and edit the result, as AI can generate
authoritative-sounding output that can be incorrect, incomplete or
biased. AI and AI-assisted technologies should not be listed as an
author or co-author, or be cited as an author. Authorship implies
responsibilities and tasks that can only be attributed to and performed
by humans, as outlined in Elsevier's AI policy for authors.

Authors should disclose in their manuscript the use of AI and
AI-assisted technologies in the writing process by following the
instructions below. A statement will appear in the published work.
Please note that authors are ultimately responsible and accountable for
the contents of the work.

Authors must disclose the use of generative AI and AI-assisted
technologies in the writing process by adding a statement at the end of
their manuscript in the core manuscript file, before the References
list. The statement should be placed in a new section entitled
`Declaration of Generative AI and AI-assisted technologies in the
writing process'.

Statement: During the preparation of this work the author(s) used
{[}NAME TOOL / SERVICE{]} in order to {[}REASON{]}. After using this
tool/service, the author(s) reviewed and edited the content as needed
and take(s) full responsibility for the content of the publication.

This declaration does not apply to the use of basic tools for checking
grammar, spelling, references etc. If there is nothing to disclose,
there is no need to add a statement.
\end{quote}

\begin{enumerate}
\def\labelenumi{\arabic{enumi}.}
\setcounter{enumi}{6}
\item
  \begin{quote}
  \textbf{Asian-Australasian Journal of Animal Sciences}: N/A
  \end{quote}
\item
  \begin{quote}
  \textbf{Theriogenology}: (Same as in ``Veterinary Microbiology'')
  \end{quote}
\item
  \begin{quote}
  \textbf{Ticks and Tick-borne Diseases}: (Same as in ``Veterinary
  Microbiology'')
  \end{quote}
\item
  \begin{quote}
  \textbf{Animal Nutrition}: N/A
  \end{quote}
\item
  \begin{quote}
  \textbf{Research in Veterinary Science}: (Same as in ``Veterinary
  Microbiology'')
  \end{quote}
\item
  \begin{quote}
  \textbf{Veterinary World}: N/A
  \end{quote}
\item
  \begin{quote}
  \textbf{Preventive Veterinary Medicine}: (Same as in ``Veterinary
  Microbiology'')
  \end{quote}
\item
  \begin{quote}
  \textbf{Applied Animal Behaviour Science}: (Same as in ``Veterinary
  Microbiology'')
  \end{quote}
\item
  \begin{quote}
  \textbf{Veterinary Parasitology}: (Same as in ``Veterinary
  Microbiology'')
  \end{quote}
\item
  \begin{quote}
  \textbf{Annual Review of Animal Biosciences}: N/A
  \end{quote}
\item
  \begin{quote}
  \textbf{Tropical Animal Health and Production}: N/A
  \end{quote}
\item
  \begin{quote}
  \textbf{The Veterinary Journal}: (Same as in ``Veterinary
  Microbiology'')
  \end{quote}
\item
  \begin{quote}
  \textbf{Veterinary Research}: (Same as in ``BMC Veterinary Research'')
  \end{quote}
\item
  \begin{quote}
  \textbf{Veterinary Clinics: Food Animal Practice}:
  \end{quote}
\end{enumerate}

\begin{quote}
Where authors use artificial intelligence (AI) and AI-assisted
technologies in the writing process, authors should:

Only use these technologies to improve readability and language, not to
replace key researcher tasks such as interpreting data or drawing
scientific conclusions.

Apply the technology with human oversight and control, and carefully
review and edit the result, as AI can generate authoritative-sounding
output that can be incorrect, incomplete or biased.

Not list AI and AI-assisted technologies as an author or co-author, or
cite AI as an author. Authorship implies responsibilities and tasks that
can only be attributed to and performed by humans, as outlined in
Elsevier's AI policy for authors.

Disclose in their manuscript the use of AI and AI-assisted technologies
in the writing process by following the instructions below. A statement
will appear in the published work. Please note that authors are
ultimately responsible and accountable for the contents of the work.
\end{quote}

\textbf{ChatGPT's Ethical Issues and Limitations}

\emph{Hallucination}

\textbf{Prompt:} Can you cite a few studies where ChatGPT was used in
veterinary clinics to diagnose canine diseases, specifying the journal
names and publication years, and PMID?

\textbf{ChatGPT 4:}

Based on the information gathered, it appears that studies involving
ChatGPT in veterinary medicine are still emerging, and direct references
to specific studies with journal names, publication years, and PMIDs are
limited in the provided sources. The articles from Frontiers in
Veterinary Science and research on ChatGPT for veterinary anatomy
education from ResearchGate highlight the application and potential of
ChatGPT in veterinary medicine but do not specify journal details or
PMIDs for these studies.

The first source discusses the use of ChatGPT in diagnosing idiopathic
epilepsy in dogs, indicating the model\textquotesingle s diagnostic
capabilities can be improved with the inclusion of clinical and
neurological examination information. It also raises concerns about the
lack of scientific literature support for ChatGPT\textquotesingle s
generated content and the influence of subtle changes in medical records
on its diagnostic performance\hspace{0pt}\hspace{0pt}
{[}\href{https://www.frontiersin.org/articles/10.3389/fvets.2023.1245168/full}{``}{]}.

The second source provides an overview of ChatGPT\textquotesingle s
application in veterinary anatomy education, showcasing its potential to
describe anatomical features and procedures, such as the forelimb bones
of an ox and the site for medial patellar desmotomy in large ruminants.
This example indicates ChatGPT\textquotesingle s utility in educational
contexts for veterinary studies\hspace{0pt}\hspace{0pt}
{[}\href{https://www.researchgate.net/publication/372246495_ChatGPT_for_veterinary_anatomy_education_an_overview_of_the_prospects_and_drawbacks}{``}{]}.

While these sources illustrate ChatGPT\textquotesingle s applications in
veterinary medicine, they do not provide specific details about
published studies in peer-reviewed journals, including publication years
and PMIDs. This might suggest that the field is still in the early
stages of exploring and documenting the use of AI models like ChatGPT in
veterinary research and education. Further research and publications are
likely needed to provide a comprehensive list of studies with detailed
bibliographic information.

\textbf{\hfill\break
}

\emph{Practical Learning Resources}

\textbf{Prompt:} I\textquotesingle m a veterinarian with no background
in programming. I\textquotesingle m interested in learning how to use
generative AI tools like ChatGPT. Can you recommend some resources for
beginners?

\textbf{ChatGPT 4:}

Certainly! Learning to use generative AI tools like ChatGPT can be an
exciting and rewarding journey, even for those without a background in
programming. Here are some resources tailored for beginners that can
help you get started:

\begin{enumerate}
\def\labelenumi{\arabic{enumi}.}
\item
  \textbf{OpenAI\textquotesingle s Documentation and Guides}:
\end{enumerate}

\begin{quote}
• Start with OpenAI\textquotesingle s official documentation. It
provides a comprehensive overview of how to use their models, including
ChatGPT. The guides are structured to help non-technical users
understand the basics of interacting with the model.

• URL: \href{https://help.openai.com/en/}{OpenAI Documentation}
\end{quote}

\begin{enumerate}
\def\labelenumi{\arabic{enumi}.}
\setcounter{enumi}{1}
\item
  \textbf{Online Courses}:
\end{enumerate}

\begin{quote}
• There are online platforms like Coursera, Udemy, and edX that offer
introductory courses on AI and machine learning. Look for courses that
focus on practical applications of AI or specifically on generative
models.

• Some courses may be titled "Introduction to AI" or "Machine Learning
for Everyone" and can provide a good foundation without requiring a
programming background.
\end{quote}

\begin{enumerate}
\def\labelenumi{\arabic{enumi}.}
\setcounter{enumi}{2}
\item
  \textbf{YouTube Tutorials}:
\end{enumerate}

\begin{quote}
• YouTube is a great resource for visual learners. Channels like "Two
Minute Papers," "CodeEmporium," and "Siraj Raval" break down complex AI
concepts into more digestible content. You can find tutorials
specifically about generative AI models and their applications.
\end{quote}

\begin{enumerate}
\def\labelenumi{\arabic{enumi}.}
\setcounter{enumi}{3}
\item
  \textbf{Blogs and Articles}:
\end{enumerate}

\begin{quote}
• Many tech blogs and websites publish articles aimed at beginners.
Websites like Medium, Towards Data Science, and Analytics Vidhya have
user-friendly articles that explain AI concepts and tools in simple
terms.
\end{quote}

\begin{enumerate}
\def\labelenumi{\arabic{enumi}.}
\setcounter{enumi}{4}
\item
  \textbf{Communities and Forums}:
\end{enumerate}

\begin{quote}
• Joining communities like Reddit (e.g., r/MachineLearning,
r/LanguageTechnology) or Stack Overflow can be beneficial. You can ask
questions, share your learning progress, and connect with others who are
also learning about AI.
\end{quote}

\begin{enumerate}
\def\labelenumi{\arabic{enumi}.}
\setcounter{enumi}{5}
\item
  \textbf{Books}:
\end{enumerate}

\begin{quote}
• For those who prefer a more structured approach to learning, there are
several beginner-friendly books on AI and machine learning. Titles like
"AI for People and Business" by Alex Castrounis or "Artificial
Intelligence: A Guide for Thinking Humans" by Melanie Mitchell offer
insights into AI without the need for a technical background.
\end{quote}

\begin{enumerate}
\def\labelenumi{\arabic{enumi}.}
\setcounter{enumi}{6}
\item
  \textbf{Workshops and Webinars}:
\end{enumerate}

\begin{quote}
• Keep an eye out for workshops, webinars, and public talks about AI and
generative models. Universities, tech companies, and professional
organizations often host events that are accessible to beginners.
\end{quote}